\documentclass{mem}
\usepackage{natbib}\usepackage{txfonts}\usepackage{balance}
\usepackage{graphicx}
\usepackage[a4paper,breaklinks,dvipdfm]{hyperref}
\idline{0}{0}
\begin{document}

\title{Examining the Age/Activity Relationship of Ultracool Dwarfs with GAIA }

\subtitle{}

\author{S.\ J.\ Schmidt \inst{1} }

\offprints{S. J. Schmidt}
 
\institute{
The Ohio State University,
140 West 18th Ave.,
Columbus, OH 43210; 
\email{schmidt@astronomy.ohio-state.edu}
}

\authorrunning{Schmidt}

\titlerunning{Ultracool Dwarf Age/Activity}

\abstract{
The relationship between age, rotation, and magnetic activity can be used to roughly estimate the ages of solar-type stars. At lower stellar masses, the relationship between activity and age changes due to the less efficient angular momentum loss, and may disappear entirely for ultracool (late-M and L) dwarfs. The detection of flares (as a tracer of magnetic activity) can be combined with kinematic tracers of age to explore the relationship between age and activity for ultracool dwarfs. The final data release of GAIA will provide time-resolved photometry of GAIA targets in the $G$ filter, but the effect of flares in the $G$-band is not well understood. I use a simple flare model to estimate the conversion of flare magnitudes for M3--L5 dwarfs in the Johnson $V$-, SDSS $r$-, and Kepler- bands to the GAIA $G$-band. By applying those conversions to previously observed flare rates, I estimate that M0-M6 dwarfs will have a flare rate in GAIA of $R_{f,GAIA} \gtrsim 8.73 \times 10^{-4}~{\rm hr}^{-1}~{\rm deg}^{-2}$, corresponding to a total of $\gtrsim$20,000 flares in the whole survey. If ultracool dwarfs have the same flare rate, I would expect a total of $\gtrsim$20 flares on M7-L5 dwarfs. 
}
\maketitle{}

\section{Introduction}
``Activity" is a blanket term that describes any interaction of surface magnetic fields with the material at and above the photosphere of a star or brown dwarf. Tracers of activity are either quisi-static/quiescent (e.g., H$\alpha$ or Ca II H\&K, which trace heated regions  likely associated with starspots) or are part of flares (dramatic events that result in emission across the electromagnetic spectrum). As stars age, they dissipate angular momentum and their rotation gradually slows. Fast rotators have stronger magnetic fields than slow rotators, so overall magnetic field strength declines with age. As the magnetic field abates, activity indicators weaken or disappear because the surface magnetic field interacts less with the stellar surface. 

This overall relationship between age and activity changes with stellar mass. The Ca II H\&K index of solar-type stars declines with age, allowing rough age-dating of younger FGK stars \citep[e.g.,][]{Mamajek2008}. Early- to mid- M dwarfs (spectral types M0--M7) are characterized by activity lifetimes---H$\alpha$ is in emission for 1--8~Gyrs (1--2~Gyr for M0--M2 dwarfs, increasing to 8~Gyr for M7 dwarfs), then the star becomes inactive \citep{West2008}. The relationship between activity and age for ultracool dwarfs is unclear; observations of H$\alpha$ emission compared to Galactic height indicate that late-M and L dwarfs may follow a similar activity lifetime relationship to the earlier-M dwarfs, but with longer active lifetimes \citep{Schmidt2012phd}. Observations of rotation in ultracool dwarfs indicate a breakdown of the age/rotation relationship for L dwarfs \citep{Reiners2008}. It is possible that the angular momentum loss of L dwarfs is so slow that even at 10~Gyr, they still have strong magnetic fields. 

The presence and strength of flares may prove to be a good age indicator for M and L dwarfs. Based on data from the Sloan Digital Sky Survey \citep[SDSS;][]{York2000}, both \citet{Kowalski2009} and \citet{Hilton2010} found that M0-M6 dwarfs that flare are preferentially found at lower Galactic heights than stars with H$\alpha$ emission, implying a ``flare lifetime" for M0-M6 dwarfs that is shorter than the activity lifetime. Ultracool dwarfs could follow a similar pattern, with flare lifetimes of a few Gyrs. Recently, an M8 and an L1 dwarf have both been observed during dramatic flares, and neither shows any significant signs of youth \citep{Gizis2013,Schmidt2014}. 

The GAIA mission presents an ideal opportunity to investigate the relationship between flare activity and age. While repeat observations are intended primarily to derive an astrometric solution, they can also be used to identify and classify flare stars. Those flare detections can be combined with kinematics to understand the relationship between flare activity and age. In the rest of the proceedings, I estimate the flare magnitudes that could be observed by GAIA (Section~2) and a rough flare rate for GAIA (Section~3).

\section{Estimating Flare Magnitudes in the GAIA Filter}
Flares are characterized by high temperatures, so they are best observed with relatively blue filters (e.g., the SDSS $u$- or $g$-band) where the hot, blue flare emission is much stronger than then the cool, red surface of the star. The GAIA $G$ filter is not an ideal flare filter; it is both very broad, encompassing 3200--10000~\AA, and relatively red, with a peak wavelength of $\sim6000$~\AA. To estimate the effect of flares on the $G$-band, I adapted the \citet{Davenport2012} technique. I assembled spectroscopic templates of M3--L5 dwarfs, then generated a suite of flares using a $T=10000$~K blackbody scaled to different physical extents. 

For wavelengths bluer than $\lambda <9200$~\AA, I used the M dwarf template spectra from \citet{Bochanski2007a} and the L dwarf template spectra from Schmidt et al. (2014b, submitted to PASP). The templates do not extend all the way to 3200~\AA; I set the flux to zero between the end of the template ($\lambda < 3800$~\AA~for M3--M9 and $\lambda < 5200$~\AA~for L0--L5) and 3200~\AA. This should not have a strong effect on the $G$-band flux, but prevents us from calculating comparisons between blue filters (e.g., SDSS $u$ and $g$) with the $G$-band. To extend the templates to redder wavelengths, I combined each template with infrared spectra of individual objects from the SpeX PRISM archives (listed in Table~\ref{tab:st}).

\begin{table}
\caption{Details for Each Spectral Type}
\label{tab:st}
\begin{tabular}{llllll}
\hline
ST  & \multicolumn{2}{c}{Infrared Spectrum} & \multicolumn{3}{c}{$\Delta G = 1$} \\
& Object & Ref &$\Delta V$ & $\Delta r$ & $\Delta$Kepler  \\
\hline
M3 & NLTT 57259  & 1& 1.6 & 1.2 & 0.36 \\
M4  &  LP 508-14 & 2 & 1.7 & 1.3 & 0.34 \\
M5  &   Gl 866AB & 3 & 1.9 & 1.4 & 0.33 \\
M6  &  Wolf 359 & 3 & 2.2 & 1.6 & 0.31 \\
M7  &  VB 8  & 3 & 2.6 & 1.9 & 0.31 \\
M8  & VB 10 & 2 & 3.2 & 2.5 & 0.33 \\
M9  & LHS 2924 & 4 & 3.3 & 2.5 & 0.35 \\
L0  & 2M0345+25 & 4 & 3.8 & 2.4 & 0.37 \\
L1  & 2M1439+19 & 2 & 3.9 & 2.4 & 0.36 \\
L2  & Kelu-I & 5 & 3.9 & 2.3 & 0.35 \\
L3  &  2M1146+22 & 6 & 4.1 & 2.3 & 0.36 \\
L4  & 2M1104+19 & 2 & 3.9 & 2.3 & 0.40 \\
L5  & 2M1239+55 & 6 & 4.5 & 2.5 & 0.42 \\
\hline
\end{tabular}
References: (1) \citet{Kirkpatrick2010}; (2) \citet{Burgasser2004}; (3) \citet{Burgasser2008b}; (4) \citet{Burgasser2006a}; (5) \citet{Burgasser2007}; (6) \citet{Burgasser2010} 
\end{table}

Flare emission can be modeled by a combination of Balmer emission and a thermal spectrum \citep[e.g.,][]{Kowalski2013}. Because the majority of Balmer emission occurs at wavelengths bluer than $\sim4000$~\AA (where the $G$-band response is low), I adopt a simplified model of a flare as a single $T=10000$~K blackbody spectrum. To produce different flare magnitudes, I scaled the surface coverage of the blackbody emission to produce a range of observable flares when contrasted with the quiescent spectrum. The flare response for Johnson $V$, SDSS $r$, and the Kepler band are given for $\Delta G = 1$ in Table~\ref{tab:st} and shown for a range of $\Delta G$ and spectral type in Figure~1. 

\begin{figure}[t!]
\resizebox{\hsize}{!}{\includegraphics[clip=true]{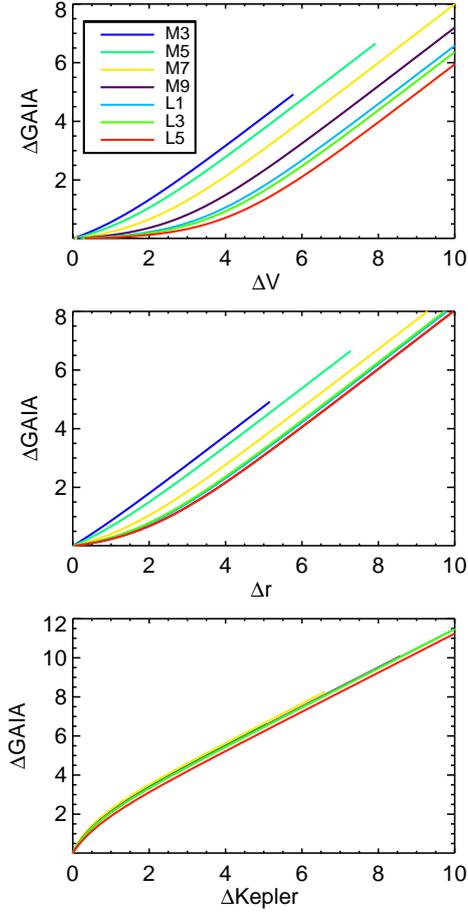}}
\caption{\footnotesize
The change in magnitude in the $G$ ($\Delta G$) due to a flare (modeled by a $T=10000$~K blackbody) as a function of the change in magnitude in Johnson $V$, SDSS $r$, and the Kepler filters.
}
\label{fig:delmag}
\end{figure}

\section{How Many UCD Flares Will GAIA Observe?}
The flare rate of ultracool dwarfs has not been thoroughly measured, but a simple estimate can be obtained by scaling the flare rate of M0-M6 dwarfs. \citet{Kowalski2009} calculated a flare rate for M0-M6 dwarfs in repeat observations of Stripe82 from SDSS, finding that flares larger than $\Delta u>0.7$ on M0-M6 dwarfs brighter than $u<22$ have a rate of  $R_{f,S82} = 1.2$~hr$^{-1}$~deg$^{-2}$. A first estimate of the flare rate in GAIA can be obtained by scaling this estimate as follows:  

\begin{equation}
R_{f,GAIA} =  \left(\frac{d_{max,GAIA}}{d_{max,S82}}\right)^3 \times f_{f,S82} \times R_{f,S82},
\end{equation}

where the rate of flares in GAIA, $R_{f,GAIA}$, is based on the ratio of the volumes observed within the distance limits of each survey (based on the maximum observed distances of GAIA, $d_{max,GAIA}$, and Stripe82, $d_{max,Stripe82}$) and $f_{f,S82}$, the fraction of Stripe82 flares that would be observable in the GAIA $G$-band.

Using the \citet{West2011} catalog of M dwarfs, I estimate a constant $u-r \sim 3.7$ over M0--M6 spectral types, and spectrophotometry indicates a roughly constant $G-r \sim 4.5$. The limit of $u < 22$ roughly corresponds to $G_{lim,S82} < 23$. To scale the flare rate to GAIA, I apply the effect of the limiting magnitude ($G_{lim,GAIA}<20$) on the volume observed:

\begin{equation} 
\left(\frac{d_{max,GAIA}}{d_{max,S82}}\right)^3 = 10^{0.6\times (G_{lim,GAIA} - G_{lim,S82})}= 0.0158.
\end{equation}

We next scale by the fraction of Stripe82 flares that will be visible in $G$-band. To scale from $\Delta u$ to $\Delta G$, I first used the \citet{Davenport2012} relation between $\Delta u$ and $\Delta r$ for M6 dwarfs, then the relationship between $\Delta r$ and $\Delta G$ for M6 shown in Figure~\ref{fig:delmag}. A $\Delta u = 0.7$ flare would be equivalent to a $\Delta G = 0.002$ flare, which is too small to be detected in individual exposures. A $\Delta u = 5$ flare, however, would correspond to a $\Delta G = 0.04$ flare, likely to be above the detection threshold for a single visit. Of the 217 flares detected in Stripe82, only one was $\Delta u > 5$, resulting in $f_{f,S82} = \frac{1}{217}$.

The scaled flare rate in GAIA is then $R_{f,GAIA} =  8.73 \times 10^{-4}~{\rm hr}^{-1}~{\rm deg}^{-2}$, significantly lower than the Stripe82 flare rate due both to the smaller number of M0-M6 dwarfs observed per square degree and to the decreased flare sensitivity of the $G$-band compared to the $u$-band. This flare rate can be used to estimate the total number of M dwarf flares observed in GAIA as follows: 

\begin{equation}
N_{f,GAIA} = {\rm \#~exp}\times ET \times Area \times R_{flare,GAIA},
\end{equation}

where the \#~exp is the average number of visits (70) and the total exposure time per observation is $ET = 79$~s (2 telescopes each with 9 ccds with individual $ET=4.4$~s). GAIA covers the whole sky ($Area=41253$~deg$^{-2}$), resulting in $\sim$20,000 flares on M0-M6 dwarfs.

There are no strong constraints on the flare rate of ultracool dwarfs; if they flare at the same rate as M dwarfs then the flare rate of ultracool dwarfs can be scaled by the ratio of M0--M6 dwarfs to M7--L3 dwarfs expected in GAIA. According to simulations \citep{Robin2012,Sarro2013}, there should be $\sim$1000 times as many M0--M6 dwarfs as M7--L5 dwarfs, resulting in an ultracool dwarf flare rate of $8.73\times 10^{-7}~{\rm hr}^{-1}~{\rm deg}^{-2}$, corresponding to a total of $\sim$20 flares.

These numbers are likely to be only lower limits on GAIA flare observations. Stripe82 covers an area around the southern Galactic pole, and its deeper magnitude limit observed M dwarfs to greater distances than GAIA will sample. Over the whole sky, GAIA will observe more M dwarfs per square degree, and on average they will be younger, resulting in an increase in observed flares.

\section{Summary}
The GAIA $G$-band is not ideal for observations of flares, but due to the large number of stars observed an all-sky coverage, the survey is likely to observe at least $\sim$20,000 flares on M0-M6 dwarfs and $\sim$20 flares on M7-L5 dwarfs. These flares can be combined with other indicators of age (e.g., kinematics) to examine the age/activity relationship of cool and ultracool dwarfs.

\begin{acknowledgements}
I thank J.\ R.\ A.\ Davenport for useful discussions, and for the attendees of the GAIA Brown Dwarfs conference for excellent feedback on flares in GAIA. This research has benefitted from the SpeX Prism Spectral Libraries, maintained by Adam Burgasser at \url{http://pono.ucsd.edu/\~adam/browndwarfs/spexprism}.
\end{acknowledgements}


\begin{thebibliography}{21}
\expandafter\ifx\csname natexlab\endcsname\relax\def\natexlab#1{#1}\fi

\bibitem[{{Bochanski} {et~al.}(2007){Bochanski}, {West}, {Hawley}, \&
  {Covey}}]{Bochanski2007a}
{Bochanski}, J.~J., {West}, A.~A., {Hawley}, S.~L., \& {Covey}, K.~R. 2007,
  \aj, 133, 531

\bibitem[{{Burgasser} {et~al.}(2010){Burgasser}, {Cruz}, {Cushing}, {Gelino},
  {Looper}, {Faherty}, {Kirkpatrick}, \& {Reid}}]{Burgasser2010}
{Burgasser}, A.~J., {Cruz}, K.~L., {Cushing}, M., {et~al.} 2010, \apj, 710,
  1142

\bibitem[{{Burgasser} {et~al.}(2007){Burgasser}, {Cruz}, \&
  {Kirkpatrick}}]{Burgasser2007}
{Burgasser}, A.~J., {Cruz}, K.~L., \& {Kirkpatrick}, J.~D. 2007, \apj, 657, 494

\bibitem[{{Burgasser} {et~al.}(2008){Burgasser}, {Liu}, {Ireland}, {Cruz}, \&
  {Dupuy}}]{Burgasser2008b}
{Burgasser}, A.~J., {Liu}, M.~C., {Ireland}, M.~J., {Cruz}, K.~L., \& {Dupuy},
  T.~J. 2008, \apj, 681, 579

\bibitem[{{Burgasser} \& {McElwain}(2006)}]{Burgasser2006a}
{Burgasser}, A.~J. \& {McElwain}, M.~W. 2006, \aj, 131, 1007

\bibitem[{{Burgasser} {et~al.}(2004){Burgasser}, {McElwain}, {Kirkpatrick},
  {Cruz}, {Tinney}, \& {Reid}}]{Burgasser2004}
{Burgasser}, A.~J., {McElwain}, M.~W., {Kirkpatrick}, J.~D., {et~al.} 2004,
  \aj, 127, 2856

\bibitem[{{Davenport} {et~al.}(2012){Davenport}, {Becker}, {Kowalski},
  {Hawley}, {Schmidt}, {Hilton}, {Sesar}, \& {Cutri}}]{Davenport2012}
{Davenport}, J.~R.~A., {Becker}, A.~C., {Kowalski}, A.~F., {et~al.} 2012, \apj,
  748, 58

\bibitem[{{Gizis} {et~al.}(2013){Gizis}, {Burgasser}, {Berger}, {Williams},
  {Vrba}, {Cruz}, \& {Metchev}}]{Gizis2013}
{Gizis}, J.~E., {Burgasser}, A.~J., {Berger}, E., {et~al.} 2013, \apj, 779, 172

\bibitem[{{Hilton} {et~al.}(2010){Hilton}, {West}, {Hawley}, \&
  {Kowalski}}]{Hilton2010}
{Hilton}, E.~J., {West}, A.~A., {Hawley}, S.~L., \& {Kowalski}, A.~F. 2010,
  \aj, 140, 1402

\bibitem[{{Kirkpatrick} {et~al.}(2010){Kirkpatrick}, {Looper}, {Burgasser},
  {Schurr}, {Cutri}, {Cushing}, {Cruz}, {Sweet}, {Knapp}, {Barman},
  {Bochanski}, {Roellig}, {McLean}, {McGovern}, \& {Rice}}]{Kirkpatrick2010}
{Kirkpatrick}, J.~D., {Looper}, D.~L., {Burgasser}, A.~J., {et~al.} 2010,
  \apjs, 190, 100

\bibitem[{{Kowalski} {et~al.}(2009){Kowalski}, {Hawley}, {Hilton}, {Becker},
  {West}, {Bochanski}, \& {Sesar}}]{Kowalski2009}
{Kowalski}, A.~F., {Hawley}, S.~L., {Hilton}, E.~J., {et~al.} 2009, \aj, 138,
  633

\bibitem[{{Kowalski} {et~al.}(2013){Kowalski}, {Hawley}, {Wisniewski}, {Osten},
  {Hilton}, {Holtzman}, {Schmidt}, \& {Davenport}}]{Kowalski2013}
{Kowalski}, A.~F., {Hawley}, S.~L., {Wisniewski}, J.~P., {et~al.} 2013, \apjs,
  207, 15

\bibitem[{{Mamajek} \& {Hillenbrand}(2008)}]{Mamajek2008}
{Mamajek}, E.~E. \& {Hillenbrand}, L.~A. 2008, \apj, 687, 1264

\bibitem[{{Reiners} \& {Basri}(2008)}]{Reiners2008}
{Reiners}, A. \& {Basri}, G. 2008, \apj, 684, 1390

\bibitem[{{Robin} {et~al.}(2012){Robin}, {Luri}, {Reyl{\'e}}, {Isasi}, {Grux},
  {Blanco-Cuaresma}, {Arenou}, {Babusiaux}, {Belcheva}, {Drimmel}, {Jordi},
  {Krone-Martins}, {Masana}, {Mauduit}, {Mignard}, {Mowlavi},
  {Rocca-Volmerange}, {Sartoretti}, {Slezak}, \& {Sozzetti}}]{Robin2012}
{Robin}, A.~C., {Luri}, X., {Reyl{\'e}}, C., {et~al.} 2012, \aap, 543, A100

\bibitem[{{Sarro} {et~al.}(2013){Sarro}, {Berihuete}, {Carri{\'o}n}, {Barrado},
  {Cruz}, \& {Isasi}}]{Sarro2013}
{Sarro}, L.~M., {Berihuete}, A., {Carri{\'o}n}, C., {et~al.} 2013, \aap, 550,
  A44

\bibitem[{{Schmidt}(2012)}]{Schmidt2012phd}
{Schmidt}, S.~J. 2012, PhD thesis, University of Washington

\bibitem[{{Schmidt} {et~al.}(2014){Schmidt}, {Prieto}, {Stanek}, {Shappee},
  {Morrell}, {Bardalez Gagliuffi}, {Kochanek}, {Jencson}, {Holoien}, {Basu},
  {Beacom}, {Szczygie{\l}}, {Pojmanski}, {Brimacombe}, {Dubberley}, {Elphick},
  {Foale}, {Hawkins}, {Mullins}, {Rosing}, {Ross}, \& {Walker}}]{Schmidt2014}
{Schmidt}, S.~J., {Prieto}, J.~L., {Stanek}, K.~Z., {et~al.} 2014, \apjl, 781,
  L24

\bibitem[{{West} {et~al.}(2008){West}, {Hawley}, {Bochanski}, {Covey}, {Reid},
  {Dhital}, {Hilton}, \& {Masuda}}]{West2008}
{West}, A.~A., {Hawley}, S.~L., {Bochanski}, J.~J., {et~al.} 2008, \aj, 135,
  785

\bibitem[{{West} {et~al.}(2011){West}, {Morgan}, {Bochanski}, {Andersen},
  {Bell}, {Kowalski}, {Davenport}, {Hawley}, {Schmidt}, {Bernat}, {Hilton},
  {Muirhead}, {Covey}, {Rojas-Ayala}, {Schlawin}, {Gooding}, {Schluns},
  {Dhital}, {Pineda}, \& {Jones}}]{West2011}
{West}, A.~A., {Morgan}, D.~P., {Bochanski}, J.~J., {et~al.} 2011, \aj, 141, 97

\bibitem[{{York} {et~al.}(2000){York}, {Adelman}, {Anderson}, {Anderson},
  {Annis}, {Bahcall}, {Bakken}, {Barkhouser}, {Bastian}, {Berman}, {Boroski},
  {Bracker}, {Briegel}, {Briggs}, {Brinkmann}, {Brunner}, {Burles}, {Carey},
  {Carr}, {Castander}, {Chen}, {Colestock}, {Connolly}, {Crocker}, {Csabai},
  {Czarapata}, {Davis}, {Doi}, {Dombeck}, {Eisenstein}, {Ellman}, {Elms},
  {Evans}, {Fan}, {Federwitz}, {Fiscelli}, {Friedman}, {Frieman}, {Fukugita},
  {Gillespie}, {Gunn}, {Gurbani}, {de Haas}, {Haldeman}, {Harris}, {Hayes},
  {Heckman}, {Hennessy}, {Hindsley}, {Holm}, {Holmgren}, {Huang}, {Hull},
  {Husby}, {Ichikawa}, {Ichikawa}, {Ivezi{\'c}}, {Kent}, {Kim}, {Kinney},
  {Klaene}, {Kleinman}, {Kleinman}, {Knapp}, {Korienek}, {Kron}, {Kunszt},
  {Lamb}, {Lee}, {Leger}, {Limmongkol}, {Lindenmeyer}, {Long}, {Loomis},
  {Loveday}, {Lucinio}, {Lupton}, {MacKinnon}, {Mannery}, {Mantsch}, {Margon},
  {McGehee}, {McKay}, {Meiksin}, {Merelli}, {Monet}, {Munn}, {Narayanan},
  {Nash}, {Neilsen}, {Neswold}, {Newberg}, {Nichol}, {Nicinski}, {Nonino},
  {Okada}, {Okamura}, {Ostriker}, {Owen}, {Pauls}, {Peoples}, {Peterson},
  {Petravick}, {Pier}, {Pope}, {Pordes}, {Prosapio}, {Rechenmacher}, {Quinn},
  {Richards}, {Richmond}, {Rivetta}, {Rockosi}, {Ruthmansdorfer}, {Sandford},
  {Schlegel}, {Schneider}, {Sekiguchi}, {Sergey}, {Shimasaku}, {Siegmund},
  {Smee}, {Smith}, {Snedden}, {Stone}, {Stoughton}, {Strauss}, {Stubbs},
  {SubbaRao}, {Szalay}, {Szapudi}, {Szokoly}, {Thakar}, {Tremonti}, {Tucker},
  {Uomoto}, {Vanden Berk}, {Vogeley}, {Waddell}, {Wang}, {Watanabe},
  {Weinberg}, {Yanny}, \& {Yasuda}}]{York2000}
{York}, D.~G., {Adelman}, J., {Anderson}, Jr., J.~E., {et~al.} 2000, \aj, 120,
  1579

\end{thebibliography}
\end{document}